\documentclass[amsmath,amssymb,reprint,pre]{revtex4-2}
\PassOptionsToPackage{hyphens}{url}\usepackage{hyperref}
\hypersetup{colorlinks=false, pdfborder={0 0 0}}

\usepackage[title]{appendix}

\usepackage{amssymb,bm,mathtools,xcolor}
\usepackage{graphicx}

\DeclarePairedDelimiter{\set}{\{}{\}}

\newcommand{\eps}{\varepsilon}

\let\boldsymbol\mathbf

\def\bzero{{\boldsymbol{0}}}

\def\be{{\boldsymbol{e}}}

\def\bn{{\boldsymbol{n}}}

\def\bu{{\boldsymbol{u}}}
\def\bv{{\boldsymbol{v}}}

\def\bx{{\boldsymbol{x}}}

\def\bA{{\boldsymbol{A}}}

\def\bG{{\boldsymbol{G}}}

\def\bI{{\boldsymbol{I}}}

\def\bK{{\boldsymbol{K}}}

\def\bX{{\boldsymbol{X}}}

\def\betav{\bm{\eta}}

\def\bxi{\bm{\xi}}

\def\bbR{\mathbb{R}}
\def\bbT{\mathbb{T}}

\newcommand{\grad}{\nabla}

\renewcommand{\d}{{\rm d}}
\renewcommand{\i}{{\rm i}}

\begin{document}

\title{Geometric regulation of stress-limited growth on surfaces}
\author{Scott Weady}
\email{sweady@flatironinstitute.org}
\affiliation{Center for Computational Biology, Flatiron Institute, New York, New York 10010, USA}

\begin{abstract}
We analyze a model for microbial communities growing on surfaces in which individual growth is inhibited by collective growth-induced pressure. Asymptotic and numerical solutions on the sphere and homogenization of periodic surfaces show surface geometry can be tuned to either slow or accelerate growth. We use these insights to identify classes of surfaces for which the growth rate can be made arbitrarily small or arbitrarily close to the stress-free growth rate. These results show stress-sensing is coupled to surface curvature and topography at scales much larger than the individual, and provide design principles for controlling the growth of stress-sensitive microbial communities.
\end{abstract}

\maketitle

\section{Introduction}

Microbial communities are typically found in surface-associated states called biofilms.  Biofilms play an important role in many industrial, medical, and agricultural contexts, where they can corrode and foul materials, decay teeth and infect implants, or protect crops \cite{costerton1987bacterial, kolenbrander2000oral, fejerskov2008pathology, emmert1999biocontrol}. In such contexts, the surfaces on which biofilms grow are rarely flat. Teeth are patterned with micron-scale horizontal ridges (perikymata) and millimeter-scale mamelons and lobes that can harbor bacteria \cite{hillson2023dental}. Water-resistant plants, such as the lotus leaf, also exhibit hierarchical roughness from the nano to micron scales that inhibits initial colonization \cite{neinhuis1997characterization,chung2007impact,zhang2017designing,jiang2020lotus,hall2021bacterial,pellegrino2022microbial}.

Laboratory studies show such surface features can significantly influence biofilm development. Topography can influence the direction of growth, microstructural configurations, and genetic mixing \cite{gu2016escherichia,postek2024substrate}. Moreover, topography affects nutrient accessibility which generates preferential regions of growth \cite{chang2015biofilm}. Surface rheology and the associated deformations induced by growth can also affect macroscopic growth rates, which tend to be positively correlated with the shear modulus \cite{asp2022spreading,faiza2025substrate}. Such properties also dramatically affect community morphology and stability \cite{atis2019microbial,fei2020nonuniform,gonzalez2025morphogenesis,herrera2026instability}. 

Many mathematical models for biofilm development have been proposed and naturally range in complexity \cite{klapper2010mathematical,wilking2011biofilms,allen2018bacterial,hallatschek2023proliferating,porter2025growth}. Active fluid models in particular are a popular class in which the velocity of the biofilm and its associated order parameters are driven by active stresses and/or mass sources. These models often admit analytical and numerical solutions, and successfully predict large-scale morphology, interfacial instabilities, and interior pattern formation \cite{you2018geometry,martinez2022morphological,weady2024mechanics,martinez2025interfacial}. 

In recent contexts, mechanosensing, or how individuals sense and respond to the mechanics of their surroundings, has been recognized as an important factor in microbial growth \cite{belas2014biofilms,dufrene2020mechanomicrobiology,wittmann2023collective,ye2024fingering,Rahbar2025}. Growth-induced stresses in particular are important for biofilm formation as they can trigger production of the extracellular matrix \cite{Chu2018} and inhibit growth \cite{Delarue2016}. In this work, we study theoretically how mechanosensing, specifically growth that slows under growth-induced pressure, affects the expansion of microbial communities on surfaces. We consider one of the simplest free boundary models for microbial growth in which the cell velocity field is a potential subject to a volumetric growth constraint \cite{klapper2010mathematical, wang2010review,friedman2015free}. Using a combination of analytical methods and numerical simulations, we show stress-sensing couples to surface curvature and topography at scales much larger than the individual. These results consequently motivate geometric design mechanisms for controlling the spread of stress-sensing microbial communities. 

The paper is organized as follows. In Section \ref{sec:model} we introduce the mathematical model and discuss its analytical properties. In Section \ref{sec:sphere} we study growth on the sphere which provides insight into the coupling between stress-sensing and geometry. We then consider growth on periodic surfaces in Section \ref{sec:rough}, which we analyze using homogenization. This analysis shows surface anisotropy can be used to accelerate growth. These particular classes of surfaces motivate the question of the existence of surfaces which maximize or minimize growth for a given surface area constraint, which we address in Section \ref{sec:optimization}. 

\section{Mathematical model}\label{sec:model}

\begin{figure}[t]
    \centering
    \includegraphics[width=\linewidth, trim={0.5in 0.1in 0.5in 0}, clip]{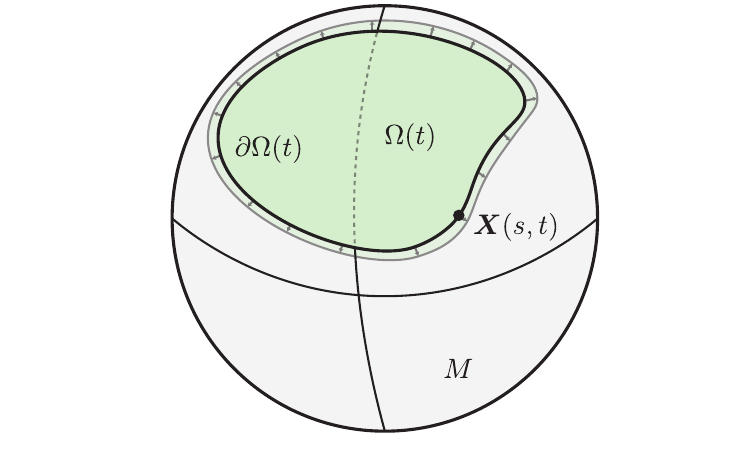}\vspace{0.125in}
    \caption{Schematic of the mathematical model. The biofilm $\Omega(t)$ (green) develops on a smooth surface $M$ (gray), which is embedded in $\bbR^3$. The evolving boundary $\partial\Omega(t)$ is parameterized by a curve $\bX : S^1 \rightarrow \bbR^3$.}\label{fig:schematic}
\end{figure}

In this section we introduce our mathematical model for a microbial community growing on a smooth surface. Throughout we refer to the microbial community as a biofilm, though we note our model does not explicitly take into account rheology of the extracellular matrix typically associated with such communities. To isolate the effects of stress response, we assume nutrients are sufficiently rich such that they do not enter the growth dynamics on the timescales considered. 

\subsection{Constitutive equations}

Let $M$ be a smooth surface and let $\Omega(t) \subset M$ be the biofilm domain, assumed to be connected with smooth boundary $\partial\Omega(t)$. We model the biofilm as an isotropic inviscid fluid with stress tensor $-p\bI$ growing on a frictional substrate, where $p : \Omega \rightarrow \bbR$ is the growth pressure. At each point in time, the cell velocity field $\bu(\bx) : \Omega \rightarrow T_{\bx}M$ satisfies a Darcy-type equation \cite{klapper2010mathematical,weady2024mechanics},
\begin{alignat}{2}
    \zeta \bu + \grad_M p &= 0 &&\quad {\rm in} ~ \Omega,\label{eq:u_dim}\\
    \nabla_M\cdot\bu &= \gamma &&\quad {\rm in} ~ \Omega,\label{eq:divu_dim}\\
    p &= 0 &&\quad {\rm on} ~ \partial\Omega,\label{eq:pbc_dim}
\end{alignat}
where $\grad_M$ and $\nabla_M\cdot$ are the surface gradient and divergence operators, respectively. Equation (\ref{eq:u_dim}) describes the inertialess force balance within the biofilm, in which gradients in the growth pressure are balanced by frictional forces with constant friction coefficient $\zeta$. These frictional forces may be interpreted as the combined effects of the surface and the extracellular matrix. Equation (\ref{eq:divu_dim}) describes local volumetric expansion with rate $\gamma$, which itself may depend on $p$, while Eq. (\ref{eq:pbc_dim}) says the boundary is stress-free. Note that $p$ may also be viewed as an effective Lagrange multiplier associated with the divergence constraint, for which $p = 0$ on $\partial\Omega$ is the natural boundary condition.

Due to growth, the biofilm domain changes in time. This evolution is described by the kinematic boundary condition
\begin{equation}
    \frac{\partial\bX}{\partial t} = \bu(\bX),\label{eq:dX/dt-dim}
\end{equation}
where $\bX : S^1\rightarrow \partial\Omega$ is a parameterization of the boundary. In principle only the normal velocity needs to be prescribed, though this form is consistent with spatial characteristics of the microscopic formulation of Eqs. (\ref{eq:u_dim})-(\ref{eq:pbc_dim}) \cite{weady2024mechanics}. From the kinematic condition (\ref{eq:dX/dt-dim}), this implies the surface area $|\Omega| = \int_\Omega \d S$ of the biofilm evolves as
\begin{equation}
|\dot\Omega| = \int_{\partial \Omega} \bu\cdot\hat\bn ~ \d \Gamma = \int_{\Omega} \gamma ~ \d S,\label{eq:dOmega/dt}
\end{equation}
where $\d \Gamma$ is the arclength measure, $\d S$ is the surface measure, and $\hat\bn$ is the normal to the boundary, which is tangent to the surface. 

The growth function $\gamma$ is the primary source of modeling and may depend on a variety of properties such as nutrient concentration, multispecies interactions, or substrate rheology. When $\gamma = \gamma(p)$ depends on the growth-induced pressure specifically, we say the biofilm is stress-sensing. Once such a function is specified, the set of equations (\ref{eq:u_dim})-(\ref{eq:dX/dt-dim}) is closed. In the remainder of this work we will assume $\gamma = \gamma(p)$.

It is often mathematically convenient to eliminate the velocity in Eqs. (\ref{eq:u_dim})-(\ref{eq:pbc_dim}). Taking the surface divergence of (\ref{eq:divu_dim}) gives a closed equation for the pressure,
\begin{alignat}{2}
    -\zeta^{-1}\Delta_{M} p &= \gamma(p) &&\quad {\rm in} ~ \Omega,\label{eq:Lp-dim}\\
    p &= 0 &&\quad {\rm on} ~ \partial\Omega,\label{eq:Lp-dim_bc}
\end{alignat}
where $\Delta_M$ is the Laplace-Beltrami operator. The velocity, and hence the boundary motion, can then be recovered from the pressure via Eq. (\ref{eq:u_dim}).

\subsection{Sensing slows growth}\label{sec:sensing-slows-growth}

For stress-inhibiting growth laws, the condition ${\nabla_M\cdot\bu = \gamma(p)}$ shows growth at a given point decreases as $p$ increases. However, this does not {\em a priori} imply growth of the whole biofilm will slow owing to the global coupling of the pressure. Here we show this is indeed the case for a general class of growth functions. Specifically, we assume the growth function can be expressed as $\gamma(p) = \widehat \gamma(\lambda p)$ for some non-negative parameter $\lambda$ and smooth, non-increasing, positive function $\widehat \gamma$. Under these assumptions, because $\Omega$ is assumed to be smooth and connected, $p$ is smooth, superharmonic, and therefore satisfies the strong maximum principle \cite{dacorogna2007direct,evans2022partial}, which we repeatedly invoke in the following.

Now suppose $p_1, p_2$ are solutions to (\ref{eq:Lp-dim})-(\ref{eq:Lp-dim_bc}) corresponding to parameters $\lambda_1, \lambda_2$, respectively, with $\lambda_1 > \lambda_2$. We claim $\lambda_1p_1 \geq \lambda_2 p_2$. To show this, for sake of contradiction suppose that $q := \lambda_1 p_1 - \lambda_2 p_2$ has a negative local minimum at $\bx_0 \in \Omega$, that is $q(\bx_0) < 0$ and $\Delta_{M} q(\bx_0) \geq 0$. Because $\widehat\gamma$ is non-increasing and $\lambda_1p_1(\bx_0) < \lambda_2p_2(\bx_0)$ by assumption, we must have $\widehat \gamma(\lambda_2 p_2(\bx_0)) \leq \widehat \gamma(\lambda_1p_1(\bx_0))$. Because $\widehat\gamma$ is positive and $\lambda_2 < \lambda_1$, this implies the strict inequality
\begin{equation}
    \lambda_2\widehat \gamma(\lambda_2 p_2(\bx_0)) < \lambda_1\widehat \gamma(\lambda_1 p_1(\bx_0)).
\end{equation}
However, we also have
\begin{equation}
    -\zeta^{-1}\Delta_{M}q = \lambda_1\widehat \gamma(\lambda_1 p_1) - \lambda_2\widehat \gamma(\lambda_2 p_2) \leq 0,
\end{equation}
which is a contradiction and so $\lambda_1 p_1 \geq \lambda_2p_2$.

This result has two important implications. First, since $\gamma$ is non-increasing, it shows the overall growth rate is non-increasing with respect to $\lambda$, namely, $|\dot\Omega|(\lambda_1) \leq |\dot\Omega|(\lambda_2)$. Further, since $-\Delta(p_1 - p_2) = \widehat\gamma(\lambda_1 p_1) - \widehat\gamma(\lambda_2 p_2) \leq 0$, the maximum principle shows $p_1 \leq p_2$ everywhere. These facts are surprising from a physical perspective. The latter shows pressure decreases pointwise with the sensitivity parameter $\lambda$, which we expect to accelerate growth, yet the overall growth rate of the biofilm decreases. Combined, these inequalities tightly bound the asymptotic behavior of $p$ with respect to $\lambda$ for arbitrary decreasing growth functions. Namely, if $||p||_\infty = O(\lambda^a)$, then $-1 \leq a \leq 0$.


Scaling of the growth rate for large $\lambda$ can be constrained even further by noting that $\widehat\gamma(\lambda p)$ will only be non-negligible in a region in which $\lambda p = O(1)$. By the maximum principle, this will necessarily occur in a thin region of width $\delta$ near the boundary. When $\delta$ is small compared to both the colony size and the radii of curvature of $M$ and $\partial\Omega$, we can use a boundary layer analysis to formally determine the leading order pressure gradient. Letting $\eta$ denote the geodesic distance from the boundary $\eta = 0$, to leading order the pressure satisfies $-\partial ^2p/\partial\eta^2 = \widehat\gamma(\lambda p)$ subject to the boundary conditions $p\rvert_{\eta=0} = 0$ and $\partial p/\partial\eta \to 0$ as $\eta \to \infty$. Multiplying by $\partial p/\partial \eta$ and integrating from $\eta = 0$ to $\infty$ gives
\begin{equation}
    \frac{1}{\zeta}\frac{\partial p}{\partial \eta} \Big\rvert_{\eta = 0} = \sqrt{\frac{2\Gamma}{\lambda}},
\end{equation}
where $\Gamma = \int_0^\infty \widehat\gamma(q) ~ \d q$, which is assumed to be finite. This is precisely the normal velocity, so we obtain $|\dot\Omega| \approx (2\Gamma/\lambda)^{1/2} |\partial\Omega|$, with an error associated with the thickness $\delta$. This is consistent with solutions for circular bacterial colonies in which it predicts linear asymptotic growth of the colony radius at long times \cite{weady2024mechanics}.

In the rest of this work we choose the exponentially decaying growth function $\gamma(p) = (1 / \tau) e^{-\lambda p}$, where $\tau$ is a growth time scale and $\lambda$ is the (non-negative) sensitivity parameter \cite{weady2024mechanics}. This specific form captures the physically motivated constraints on growth, namely positive and decreasing with stress, and we expect the qualitative results will be similar for other models satisfying these properties.

\subsection{Non-dimensionalization}

We non-dimensionalize the equations by the typical biofilm size $L$, growth time scale $\tau$, and stress scale $\zeta L^2 / \tau$. This yields the dimensionless velocity $\bu' = (\tau/L) \bu$, pressure $p' = (\tau/\zeta L^2) p$, and sensitivity $\lambda' = (\zeta L^2 / \tau)\lambda$. The surface is similarly rescaled by the characteristic length $L$, with the corresponding dimensionless surface and domain denoted by $M'$ and $\Omega'$, respectively. The equations of motion are then
\begin{alignat}{2}
    \bu' + \grad_{M'}' p' &= \bzero &&\quad {\rm in} ~ \Omega'\label{eq:u}\\
    \nabla_{M'}'\cdot\bu' &= e^{-\lambda'p'} && \quad{\rm in} ~ \Omega',\label{eq:divu}\\
    p' &= 0 &&\quad {\rm on} ~ \partial\Omega',\label{eq:pbc}
\end{alignat}
with the kinematic condition
\begin{equation}
    \frac{\partial\bX'}{\partial t'} = \bu'(\bX').\label{eq:dX/dt}
\end{equation}
The pressure equation is similarly
\begin{alignat}{2}
-\Delta_{M'} p' &= e^{-\lambda' p'} &&\quad {\rm in} ~ \Omega',\label{eq:Lp}\\
p' &= 0 &&\quad {\rm on} ~ \partial\Omega'\label{eq:Lp-bc}.
\end{alignat}
The only free parameter, in addition to the geometry, is the dimensionless sensitivity $\lambda'$ which accounts for the biofilm size, growth rate, and substrate friction. While evaluating the dimensional version of $\lambda$ requires specific experimental measurements, its dimensionless equivalent can be estimated based on typical biofilm expansion rates and sizes \cite{weady2024mechanics}. For biofilms of typical size $L = 1~{\rm mm}$, individual growth timescale $\tau = 1~ {\rm hr}$, and collective expansion rate $10~{\mu\rm m}/{\rm min}$, we estimate $\lambda' = O(1)-O(10)$. For ease of notation, in the following we assume all variables are dimensionless and omit primes.

\subsection{Differential geometry of surfaces}

We conclude this section by establishing our notation for the basic differential geometry of surfaces. Let $\bx(\bxi) : \Xi \rightarrow \bbR^3$ be a smooth embedding of $M$ in $\bbR^3$, with $\Xi \subset \bbR^2$ the parameter domain (i.e. local coordinates). In component form, we write a point in $\bbR^3$ as $\bx = (x, y, z)$ and a point in the parameter domain as $\bxi = (\xi, \eta)$. The matrix representation $\bG:\Xi \rightarrow \bbR^{2 \times 2}$ of the metric tensor is
\begin{equation}
\bG(\bxi) = 
\begin{pmatrix} 
\bx_\xi\cdot\bx_\xi & \bx_\xi\cdot\bx_\eta \\
\bx_\eta\cdot\bx_\xi & \bx_\eta\cdot\bx_\eta
\end{pmatrix},
\end{equation}
where subscripts denote partial derivatives, and the Jacobian is $J(\bxi) = \sqrt{\det\bG(\bxi)}$. In local coordinates, a tangent vector $\bv = (v^\xi, v^\eta)$ expressed in the (unnormalized) tangent basis $\bx_\xi,\bx_\eta$ at the point $\bx(\bxi_0)$ is identified with $\bv(\bxi_0) = (v^\xi \bx_\xi + v^\eta \bx_\eta)\rvert_{\bxi_0}$. The differential operators can then be written explicitly in local coordinates as
\begin{align}
    \grad_M v &= \bG^{-1}\nabla_{\bxi} v,\\
    \nabla_M\cdot\bv &= \frac{1}{J}\nabla_{\bxi}\cdot\left(J\bv\right),\\
    \Delta_{M}v &= \frac{1}{J}\nabla_{\bxi}\cdot\left(J\bG^{-1} \nabla_{\bxi} v\right),
\end{align}
where $\nabla_{\bxi} = (\partial_\xi, \partial_\eta)$ is the gradient operator with respect to the usual Euclidean metric.

\begin{figure*}[t!]
\centering
\includegraphics[scale=0.5]{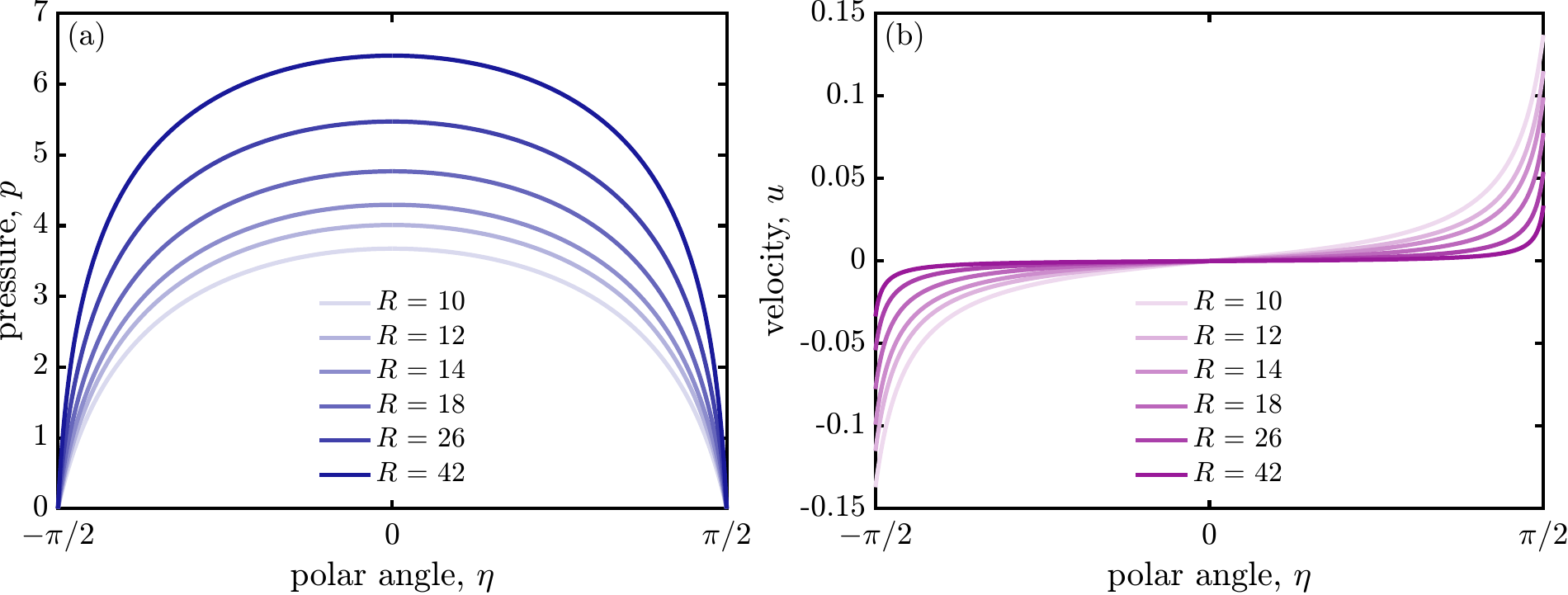}
\caption{(a) Pressure and (b) velocity profiles for growth on the sphere of radius $R$ for fixed polar angle $H = \pi/2$. The pressure increases monotonically with the radius of the sphere while the polar angle velocity decreases. Note that the area of the biofilm increases with $R$ so the comparisons in (a) and (b) are for domains with different areas. In all cases $\lambda = 1$.}\label{fig:sphere-pressure}
\end{figure*}

\section{Growth on the sphere}\label{sec:sphere}

We first consider an axisymmetric biofilm growing on a sphere. Let $M = S^2(R)$ be the sphere of radius $R$ and let
\begin{equation}
    \bx(\xi, \eta) = R\begin{pmatrix}\cos \xi \sin \eta \\ \sin \xi \sin \eta \\ \cos \eta\end{pmatrix}
\end{equation}
be a parameterization in spherical coordinates. We define the polar angle $H(t)$ such that the biofilm occupies the domain $\Omega(t) = \set{\bx(\xi, \eta) : \xi \in [0, 2\pi), \eta \in [0, H)}$. The surface area is then given by $|\Omega| = 2\pi R^2(1 - \cos H)$. Assuming axisymmetry, that is $\partial p /\partial \xi = 0$, the pressure satisfies
\begin{alignat}{2}
    -\frac{1}{R^2\sin \eta}\frac{\partial}{\partial \eta}\left(\sin \eta \frac{\partial p}{\partial \eta}\right) &= e^{-\lambda p}, &&\quad \eta \in [0,H),\label{eq:sphere-pde}\\
    \frac{\partial p}{\partial \eta} &= 0, &&\quad \eta = 0,\label{eq:sphere-bc1}\\
        p &= 0, &&\quad \eta = H.\label{eq:sphere-bc2}
\end{alignat}
In this coordinate system the kinematic condition (\ref{eq:dX/dt}) is expressed as the evolution of $H$,
\begin{equation}
    \frac{dH}{dt} = -\frac{1}{R^2} \frac{\partial p}{\partial \eta}\Big\rvert_{\eta = H}.\label{eq:sphere-dX/dt}
\end{equation}

Preliminary intuition for the connection between stress-sensing and geometry comes from a rescaling of the equations. In particular, defining $\tilde p = p/R^2$ and $\tilde\lambda = R^2\lambda$, we find
\begin{align*}
    -\frac{1}{\sin \eta}\frac{\partial}{\partial \eta}\left(\sin \eta \frac{\partial \tilde p}{\partial \eta}\right) = e^{-\tilde\lambda \tilde p}, &\quad \eta \in [0, H),\\
    \frac{dH}{dt} = -\frac{\partial\tilde p}{\partial \eta}\Big\rvert_{\eta = H},&
\end{align*}
subject to the pressure boundary conditions (\ref{eq:sphere-bc1})-(\ref{eq:sphere-bc2}). These rescaled equations are equivalent to growth on the unit sphere with a modified sensitivity $\tilde\lambda$ and thus have a few immediate implications. Naturally, for non-sensing growth, the location of the polar angle $H$ is unaffected by the radius of the sphere. Indeed, because the surface area grows exponentially, $|\Omega(t)| = |\Omega_0| e^t$, we can find the analytical expression
\begin{equation}
    H(t) = 2\sin^{-1}\left(\sin(H_0/2) e^{t/2}\right).
\end{equation}
On the other hand, for $\lambda > 0$, the effective sensitivity $\tilde\lambda$ increases quadratically with the radius of the sphere. Because the global growth rate decreases monotonically with $\lambda$, this shows the growth rate relative to the surface area of the sphere will decrease with $R$. However, this does not account for the additional scale factor of $R^2$ in the true surface area.

\begin{figure*}[t!]
\centering
\includegraphics[scale=0.5]{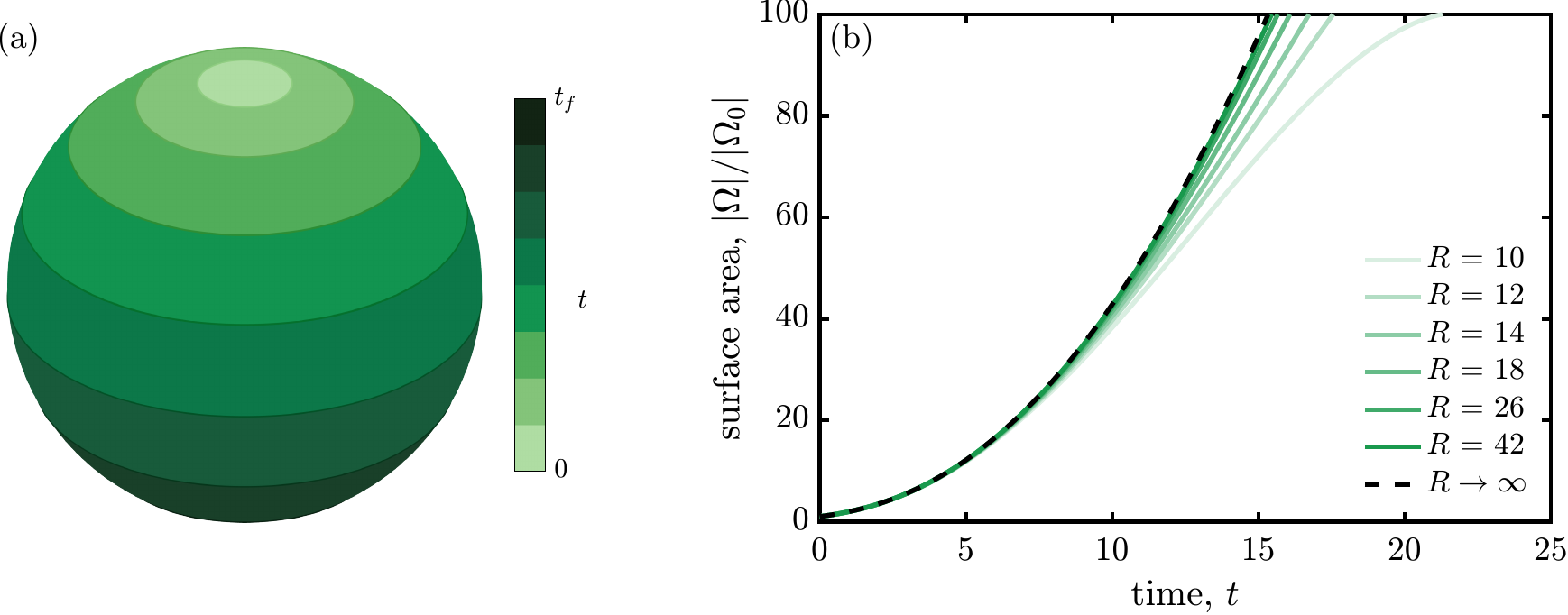}
\caption{(a) Time evolution (light to dark) of an axisymmetric biofilm growing on the sphere of radius $R = 10$. (b) Evolution of the surface area for spheres of different radii, beginning with the same initial area $|\Omega_0| = 4\pi$. The time it takes to reach a given area, here $|\Omega| = 400\pi$, decreases with $R$, approaching the planar solution as $R\rightarrow\infty$. In all cases $\lambda = 1$.}\label{fig:sphere-area}
\end{figure*}

\subsection{Asymptotic solution for large spheres}

Equations (\ref{eq:sphere-pde})-(\ref{eq:sphere-bc2}) do not readily admit an analytical solution. However, progress can be made in the infinite sphere limit, that is $R\rightarrow \infty$ with $|\Omega|$ fixed and correspondingly $H\rightarrow 0$. In particular, expanding $\cos\eta = 1 - \eta^2/2 + O(\eta^4)$, the surface area is $|\Omega| = \pi R^2 H^2 + R^2 O(H^4)$ so that convergence of the area requires $RH \rightarrow (|\Omega|/\pi)^{1/2} > 0$. This motivates the change of variables $\tilde\eta = R\eta$, for which the equations become
\begin{align*}
    -\frac{1}{\sin(\tilde\eta/R)}\frac{\partial}{\partial \tilde \eta}\left(\sin (\tilde\eta/R) \frac{\partial p}{\partial \tilde\eta}\right) = e^{-\lambda p}, &\quad \tilde\eta \in [0,  RH),\\
    \frac{d(RH)}{dt} = -\frac{\partial p}{\partial \tilde\eta}\Big\rvert_{\tilde \eta = RH},&
\end{align*}
subject again to the pressure boundary conditions (\ref{eq:sphere-bc1})-(\ref{eq:sphere-bc2}). Expanding $\sin (\tilde\eta/R) = (\tilde\eta/R) + O((\tilde\eta/R)^3)$ and defining $\tilde H = (|\Omega|/\pi)^{1/2}$, we get the asymptotic equations for $R\rightarrow\infty$,
\begin{align*}
    -\frac{1}{\tilde\eta} \frac{\partial}{\partial \tilde\eta}\left(\tilde\eta \frac{\partial p}{\partial \tilde\eta} \right) = e^{-\lambda p},&\quad \tilde\eta \in [0, \tilde H),\\
    \frac{d\tilde H}{dt} = -\frac{\partial p}{\partial \tilde\eta}\Big\rvert_{\tilde\eta = \tilde H}.&
\end{align*}
This is precisely the polar coordinate expression of Eqs. (\ref{eq:dX/dt}) and (\ref{eq:Lp}), with the radius $R$ replaced by $\tilde H$. Note that the sensitivity $\lambda$ is unchanged. The pressure and velocity can be determined analytically in this limit,
\begin{alignat}{2}
    u &= \frac{32[c(\lambda \tilde H^2)]^2\tilde \eta}{1-8[c(\lambda \tilde H^2)]^2\lambda \tilde \eta^2},&&\quad \tilde \eta \in [0, \tilde H),\label{eq:uflat}\\
     p &= \frac{2}{\lambda}\log\left(\frac{1}{8c(\lambda \tilde H^2)} - c(\lambda \tilde H^2)\lambda \tilde \eta^2\right),&&\quad \tilde \eta \in [0, \tilde H),\label{eq:pflat}
\end{alignat}
where $c(\lambda') = [(1 + \lambda'/2)^{1/2} - 1]/(2\lambda')$ \cite{weady2024mechanics}. For this planar solution, it can be shown that, for $\lambda\tilde H^2 \gg 1$, the biofilm edge grows asymptotically as $\tilde H(t) \sim (2/\lambda)^{1/2} t$, consistent with the estimation in Section \ref{sec:model}. It is also readily shown that this axisymmetric solution is linearly stable. Perturbing the boundary with wavenumber $m \geq 1$ (the case $m = 0$ is trivial) and relative amplitude $\eps(t)$, that is $\bX(\xi,t)\cdot\hat\betav(\xi) = \tilde H(1 + \eps\cos m\xi)$, we find the perturbation satisfies $\d\eps/\d t = \sigma \eps$ with
\begin{equation}
    \sigma = -4c(\lambda\tilde H^2)\,\frac{1 + 4c(\lambda\tilde H^2)(m - 1)(m + 2)}{1 + 4c(\lambda\tilde H^2)(m - 1)}.\label{eq:dispersion}
\end{equation}
(A complete derivation is provided in Appendix \ref{app:stability}). Since $c > 0$, we have $\sigma < 0$ for all $m \geq 1$ and $\lambda \geq 0$, meaning the axisymmetric solution is linearly stable. Note that, because $\tilde H$ also depends on time, this does not give $\eps(t) = \eps(0) e^{\sigma t}$, but rather monotonic decay of $\eps$.

\begin{figure*}[t!]
\centering
\includegraphics[scale=0.5]{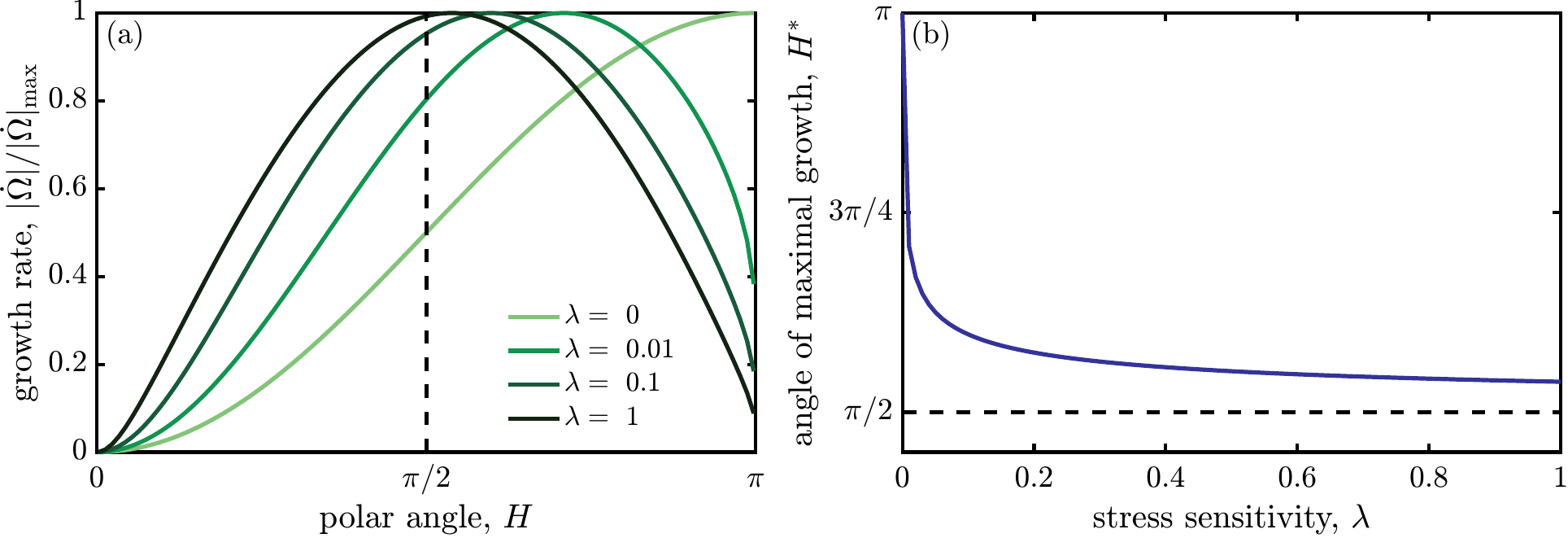}
\caption{(a) Normalized growth rate as a function of the polar angle $H$ for several values of the sensitivity parameter $\lambda$. For all positive values of $\lambda$ the growth rate exhibits a local maximum. (b) The location of the maximum $H^*$ drops sharply when $\lambda$ becomes positive and tends asymptotically to $\pi/2$, corresponding to the equator of the sphere.}\label{fig:sphere-rate}
\end{figure*}

\subsection{Numerical solution}

To assess growth on spheres of finite radii, we solve Eqs. (\ref{eq:sphere-pde})-(\ref{eq:sphere-dX/dt}) numerically. The procedure is as follows: For a given $H$, the pressure equations (\ref{eq:sphere-pde})-(\ref{eq:sphere-bc2}) are solved via a Chebyshev spectral collocation method with $N = 100$ grid points in $\eta$ along with Newton's method to handle the nonlinearity. Differential operators are constructed with \texttt{chebfun} \cite{driscoll2014chebfun}. Given the pressure, the gradient is evaluated through spectral differentiation and the interface is advanced according to Eq. (\ref{eq:sphere-dX/dt}) using a fourth-order Runge-Kutta method with time step size $\Delta t = 0.1$. 

Figure \ref{fig:sphere-pressure} shows the computed pressure and velocity profiles for fixed $H = \pi / 2$ and several values of $R$. Note that the area of the biofilm depends on $R$, so these comparisons are made for domains with different areas. The magnitude of the pressure increases while the polar velocity decreases with $R$, the latter of which is consistent with a decreasing growth rate relative to the surface area of the sphere. This tendency is contrasted by Fig. \ref{fig:sphere-area}, which shows the area over time for biofilms growing on spheres of different radii starting from the same initial surface area $|\Omega_0| = 4\pi$. At early times, the growth rate is similar across all values of $R$. However, we find the time it takes to reach a given area decreases with $R$, approaching the planar solution (black dashed line) as expected for $R\rightarrow \infty$. This shows that, among the class of spherical surfaces, the growth rate is maximized with $R\rightarrow\infty$. In particular, the time to reach $|\Omega| = 100|\Omega_0|$ for the planar case is about $25\%$ faster than for the sphere of radius $R = 10$. This behavior can be understood from the fact that, for axisymmetric biofilms of the same area, the perimeter of the domain increases with the radius of the sphere. In the absence of stress-sensing, to produce the same amount of new material, the normal velocity, and correspondingly the pressure gradient, will decrease with the radius of the sphere. This lowers pressure in the interior and we thus expect the growth rate to increase when stress-sensing is present, though we note that the nonlinear coupling does not imply this directly.

This mechanism is also evident in the smallest sphere case $R = 10$, where the growth rate $|\dot\Omega|$ is non-monotonic in the biofilm size, and the maximal growth rate occurs when the biofilm edge is near the equator $H \approx \pi/2$. Figure \ref{fig:sphere-rate} assesses this behavior in more detail. Panel (a) shows the growth rate (normalized by the maximum value for each $\lambda$) as a function of the biofilm boundary $H$. For $\lambda = 0$, the growth rate increases monotonically with $H$ as would be expected. For $\lambda > 0$, we find the growth rate exhibits a local maximum $H^* > \pi / 2$ as observed in the numerical simulations above. Moreover, the growth rate appears increasingly symmetric about $\pi / 2$. Panel (b) shows the polar angle $H^*$ of this maximal growth rate over a range of $\lambda$. We find that, consistent with the above observations, the maximum tends asymptotically to $\pi / 2$. An intuitive reason for these behaviors comes from the observation of Section \ref{sec:sensing-slows-growth} that, as $\lambda \rightarrow \infty$, growth occurs primarily at the boundary. Thus, in this limit, as the length of the boundary of the biofilm decreases so does the growth rate, though we emphasize for any finite $\lambda$ the growth rate is non-monotonic in the perimeter.

\section{Growth on periodic surfaces}\label{sec:rough}

The connection between stress-sensing and geometry identified by the solutions on the sphere motivates the introduction of geometric anisotropy through local surface fluctuations. Such patterning is observed, for example, in the lotus leaf and is known to promote hydrophobicity that can prevent initial microbial colonization \cite{neinhuis1997characterization}, though the effect of this patterning on later biofilm development is less clear. In the context of \cite{neinhuis1997characterization}, the typical periodic length scale of surface fluctuations and their height is on the order of $10~\mu\text{m}$, and biofilms in this context can be on the order of $1~\text{mm}$ in size. For cells of typical size $\sim 1 ~ \mu\text{m}$ on such a domain, the dimensionless bacterial length is then $O(10^{-3})$ while the periodic length scale $\eps$ is $O(10^{-2})$. These values are thus in a regime where the continuum model is plausible \citep{weady2024mechanics} and $\eps$ can be treated as a small parameter.

As a model for these systems, we consider surfaces periodically perturbed from a planar geometry in the form of a graph. Explicitly, we consider parameterizations of the form
\begin{equation}
\bx_\eps(\xi, \eta) = \begin{pmatrix} \xi \\ \eta \\ h_\eps(\xi, \eta)\end{pmatrix},\label{eq:x_eps}
\end{equation}
where $h_\eps(\xi,\eta) = \eps h(\xi/\eps, \eta/\eps)$ with $h(\xi', \eta')$ smooth and doubly periodic with period one. In local coordinates, the velocity and pressure $(\bu_\eps, p_\eps)$ satisfy
\begin{alignat}{2}
    \bu_\eps + \bG^{-1}_\eps\nabla_{\bxi} p_\eps &= \bzero && \quad {\rm in} ~ \Omega^*,\label{eq:u_eps}\\
    J_\eps^{-1} \nabla_{\bxi} \cdot( J_\eps \bu_\eps) &= e^{-\lambda p_\eps} && \quad {\rm in} ~ \Omega^*,\label{eq:p_eps}\\
    p_\eps &= 0 && \quad {\rm on} ~ \partial\Omega^*,\label{eq:pbc-eps}
\end{alignat}
where $\Omega^*\subset\bbR^2$ is the reference domain, $\nabla_{\bxi} = (\partial_\xi, \partial_\eta)$ denotes the gradient operator in the local coordinate system, and we've used the notation $\bu_\eps = (u_\eps^\xi, u_\eps^\eta)$. The metric tensor and Jacobian are $\bG_\eps = \bI + \nabla_{\bxi} h_\eps \otimes \nabla_{\bxi} h_\eps$ and $J_\eps = \sqrt{1 + |\nabla_{\bxi} h_\eps|^2}$. As usual, we have the pressure Poisson equation
\begin{alignat}{2}
    -\nabla_{\bxi}\cdot(J_\eps \bG^{-1}_\eps \nabla_{\bxi} p_\eps) &= J_\eps e^{-\lambda p_\eps} && \quad {\rm in} ~ \Omega^*,\label{eq:Lp-eps}\\
    p_\eps &= 0 && \quad {\rm on} ~ \partial\Omega^*.\label{eq:Lp-eps-bc}
\end{alignat}
This can be recognized as a standard homogenization problem whose asymptotic solution as $\eps\rightarrow 0$, and convergence thereof, can be found in various references \cite{allaire1992homogenization,tartar2009general}. In this section we outline this homogenization process and analyze the resulting equations. We focus on instantaneous configurations in the reference coordinate system, though in reality the interface evolution must also be homogenized.

\subsection{Homogenization}\label{sec:homogenization}

Let $\bxi' = \bxi / \eps$ be an auxiliary fast variable and define $J(\bxi') = J_\eps(\bxi)$ and $\bG(\bxi') = \bG_\eps(\bxi)$. We make a two-scale Ansatz for the pressure and velocity
\begin{align*}
p_\eps(\bxi, \bxi') &= \sum_{n = 0}^\infty p_n(\bxi, \bxi') \eps^n,\\
\bu_\eps(\bxi, \bxi') &= \sum_{n = 0}^\infty \bu_n(\bxi, \bxi')\eps^n.
\end{align*}
Substituting into Eqs. (\ref{eq:u_eps})-(\ref{eq:p_eps}) and averaging over $\bxi'$, the leading order pressure $p_0(\bxi)$ and velocity $\bu_0(\bxi)$ satisfy
\begin{alignat*}{2}
    \langle J \bu_0 \rangle + (\langle J\bG^{-1} \rangle + \bA) \nabla_{\bxi} p_0 &= \bzero \quad &&{\rm in} ~ \Omega^*,\\
    \langle J\rangle^{-1} \nabla_{\bxi}\cdot\langle J\bu_0 \rangle &= e^{-\lambda p_0} \quad &&{\rm in} ~ \Omega^*,
\end{alignat*}
where $\langle \cdot \rangle = \int_{\bbT^2} \cdot ~ \d \bxi'$ with $\bbT^2 = [0,1)^2$ the unit periodic cell. The matrix $(\bA)_{ij} = \be_i\cdot\langle J\bG^{-1}\nabla_{\bxi'} w_j \rangle$ is determined by solving periodic cell problems for auxiliary functions $w_j(\bxi')$,
\begin{equation}
    -\nabla_{\bxi'} \cdot(J\bG^{-1} \nabla_{\bxi'} w_j) = [\nabla_{\bxi'}\cdot J\bG^{-1}]\cdot\be_j \quad {\rm in} ~ \bbT^2.\label{eq:w_j}
\end{equation}
Defining $\bu^* = \langle J\bu_0 \rangle / \langle J \rangle$, $p^* = p_0$ and
\begin{equation}
\bK^* = \langle J \rangle^{-1}\left(\langle J\bG^{-1} \rangle + \bA\right)\label{eq:Kstar}
\end{equation}
gives the homogenized system of equations
\begin{alignat}{2}
\bu^* + \bK^*\nabla_{\bxi} p^* &= \bzero && \quad {\rm in} ~ \Omega^*,\label{eq:u-hom}\\
\nabla_{\bxi}\cdot\bu^* &= e^{-\lambda p^*} && \quad {\rm in} ~ \Omega^*,\label{eq:divu-hom}\\
p^* &= 0 && \quad {\rm on} ~ \partial\Omega^*.\label{eq:pbc-hom}
\end{alignat}
We refer to $\bK^*$ as the roughness tensor. Note that the surface area of the homogenized domain $\Omega^*$ is related to the true surface area by $\lim_{\eps\rightarrow0}|\Omega_\eps| = J^*|\Omega^*|$ where $J^* = \langle J \rangle$. Moreover, the homogenized velocity $\bu^*$ is not equal to $\bu_0$, which depends on the fast variable $\bxi'$, but is rather its average weighted by $J$ over the periodic cell. Relatedly, the embedded velocity $u_\eps^\xi \bx_{\eps,\xi} + u_\eps^\eta\bx_{\eps,\eta}$ will in general have a non-zero $z$-component in the limit $\eps\rightarrow0$.

\begin{figure*}[t]
    \centering
    \includegraphics[scale=0.5]{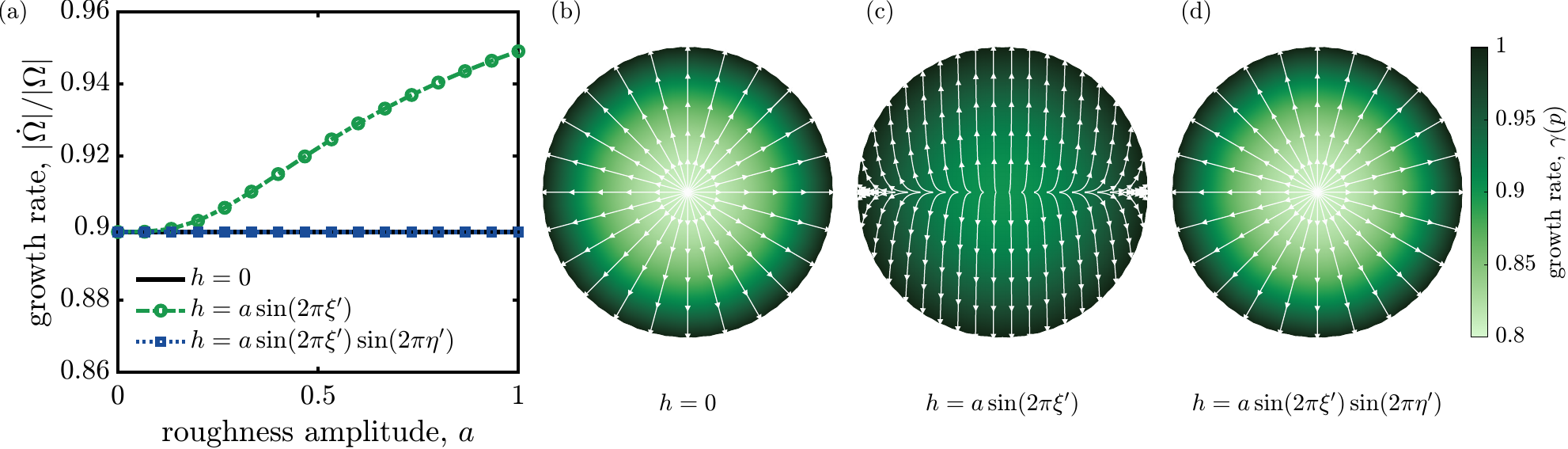}
    \caption{Growth on homogenized surfaces. Panel (a) shows the total growth rate $|\dot\Omega|/|\Omega|$ as a function of the roughness amplitude $a$ for a flat surface (black, solid), one-dimensional wave (green, dash), and two-dimensional wave (blue, dot). Panels (b)-(d) show snapshots of the pointwise growth rate along with velocity streamlines, which are radial for the flat surface and two-dimensional wave and primarily unidirectional for the one-dimensional wave. In all cases $\lambda = 1$. }
    \label{fig:homogenization}
\end{figure*}

\subsection{Bounds on the roughness tensor}

The roughness tensor $\bK^*$ is an effective mobility tensor, encoding how surface roughness biases growth and transport in different directions. This is analogous to the permeability tensor in Darcy flow, whose eigenvalues characterize the resistance to flow in the direction of the corresponding eigenvectors. In the present context, for $\mu_i(\bK^*) > \mu_j(\bK^*)$ we expect preferential growth in the direction of the $i$th eigenvector over the direction of the $j$th eigenvector. Namely, if $\mu_1 > 0$ and $\mu_2 = 0$, growth will only occur along the direction of the eigenvector associated to $\mu_1$, regardless of stress-sensing. 

The feasible values of the eigenvalues of $\bK^*$, however, are highly constrained from its relationship to the metric tensor. Specifically, because $\bG = \bI + \nabla_{\bxi} h \otimes \nabla_{\bxi} h$, the eigenvalues of $\bG$ are readily seen to be $\mu_{\rm max}(\bG) = J^2$ and $\mu_{\rm min}(\bG) = 1$ with eigenvectors $\nabla_{\bxi} h$ and $\nabla_{\bxi}^\perp h$, respectively. This observation leads to bounds on the eigenvalues of the roughness tensor $\bK^*$, namely,
\begin{equation}
\left(\frac{1}{J^*}\right)^2 \leq \mu_{\rm min}(\bK^*) \leq \mu_{\rm max}(\bK^*) \leq 1,\label{eq:mu(K)}
\end{equation}
which follows from the inequality \cite{tartar2009general}
\begin{equation}
    \langle(J\bG^{-1}) \rangle^{-1} \preceq J^*\bK^* \preceq \langle J\bG^{-1} \rangle.
\end{equation}
It is straightforward to see equality in Eq. (\ref{eq:mu(K)}) is attained only when $\bK^* = \bI$. Moreover, the lower bound can be made arbitrarily small by simply rescaling the surface function $h$.

\subsection{Anisotropy accelerates growth}

For non-sensing biofilms, the roughness tensor affects the evolving geometry but does not change the growth rate. To assess the effects of surface anisotropy in stress-sensing biofilms, here we numerically solve the homogenized equations for two example surfaces: a one-dimensional wave $h(\xi',\eta') = a\sin(2\pi\xi')$ and a two-dimensional wave $h(\xi', \eta') = a \sin (2\pi \xi') \sin (2\pi \eta')$, where $a$ is the constant roughness amplitude. While the one-dimensional wave is a developable surface and can be endowed with a Euclidean metric through a change of variables, this change of variables will modify the shape of the biofilm in reference coordinates. We thus do not expect the dynamics will be equivalent to a planar biofilm. 

To compute the roughness tensor $\bK^*$, we solve Eq. (\ref{eq:w_j}) over the unit periodic cell $\bbT^2$ using a Fourier spectral collocation method with $31 \times 31$ grid points. Given this tensor, we then solve Eqs. (\ref{eq:u-hom})-(\ref{eq:pbc-hom}) on the disk of radius $1/\sqrt{J^*}$ using a Chebyshev-Fourier spectral collocation method in polar coordinates $(r,\theta)$. We use a double cover of the disk for which $r\in(-1,1)$, taking $32$ grid points in $r$ and $31$ grid points in $\theta$. The radius is chosen such that the true surface area satisfies $|\Omega| = \pi$.

Figure \ref{fig:homogenization}(a) shows the growth rate, normalized by the surface area $|\Omega| = \pi$, for several values of the roughness amplitude $a$. For the two-dimensional wave $h(\xi',\eta') = a\sin(2\pi\xi')\sin(2\pi\eta')$, we find the growth rate is the same as that of a flat surface. In contrast, for the one-dimensional wave $h(\xi',\eta') = a\sin(2\pi\xi')$, the growth rate increases with the surface amplitude and exceeds the flat surface growth rate. This appears to approach the stress-free growth rate $\gamma\rightarrow 1$ as $a$ increases. Thus, surface anisotropy relaxes stress accumulation.

To understand this behavior, we show in panels (b)-(d) the pointwise growth rate $e^{-\lambda p^*}$ along with velocity streamlines for the flat surface, one-dimensional wave, and two-dimensional wave. The growth rate and streamlines are identical for the flat surface and the two-dimensional wave, which exhibit an axisymmetric pointwise growth rate with outward radial streamlines. The one-dimensional wave, however, exhibits a slight non-axisymmetric growth rate with streamlines that are preferentially aligned with the axis orthogonal to the wave, here $\eta$. Further, the growth rate is larger overall. 

These observations can be understood from the homogenized pressure equation, which for a diagonal roughness tensor $\bK^* = {\rm diag}(K_{11}, K_{22})$ takes the explicit form
\begin{align}
-K_{11} \frac{\partial^2 p^*}{\partial \xi^2} - K_{22} \frac{\partial^2 p^*}{\partial \eta^2} = e^{-\lambda p^*} &\quad {\rm in} ~ \Omega^* \\
p^* = 0 &\quad {\rm on} ~ \partial\Omega^*.
\end{align}
Rescaling $\tilde\xi = \xi/\sqrt{K_{11}}$ and $\tilde\eta = \eta/\sqrt{K_{22}}$, we find
\begin{align}
-\frac{\partial^2 \tilde p^*}{\partial \tilde\xi^2} - \frac{\partial^2 \tilde p^*}{\partial \tilde\eta^2} = e^{-\lambda \tilde p^*} &\quad {\rm in} ~ \tilde\Omega^* \\
\tilde p^* = 0 &\quad {\rm on} ~ \partial\tilde\Omega^*,
\end{align}
where $\tilde\Omega^* = \set{(\tilde\xi, \tilde\eta) : (\sqrt{K_{11}}\tilde\xi, \sqrt{K_{22}}\tilde\eta) \in \Omega^*}$ is an ellipse with principal axes $1/\sqrt{J^*K_{11}}$ and $1/\sqrt{J^*K_{22}}$. For our particular surfaces, numerical computations show $\bK^*$ is diagonal with $J^*\sqrt{K_{11}K_{22}} = 1$, hence the area of this ellipse is always $|\tilde\Omega^*| = \pi$. The growth rate of the physical surface is $|\dot\Omega| = \int_{\tilde\Omega^*} e^{-\lambda \tilde p^*} J^* \sqrt{K_{11} K_{22}} ~ \d S = \int_{\tilde \Omega^*} e^{-\lambda \tilde p^*} ~ \d S$, which is entirely determined by the solution on an ellipse with aspect ratio $\sqrt{K_{11}/K_{22}}$ and area $\pi$. For the two-dimensional wave, we have $K_{11} = K_{22}$ so that $\tilde\Omega^*$ is the unit disk and the growth rate is unchanged. However, for the one-dimensional wave we have $K_{11} < K_{22}$ with $K_{11}\rightarrow 0$ as $a\rightarrow\infty$ and so the domain approaches an infinitely long, infinitesimally thin line. Asymptotically, as we show in the following section, this leads to a pressure that approaches zero everywhere and a growth rate that approaches the stress-free limit.

\begin{figure*}
    \centering
    \includegraphics[width=\textwidth]{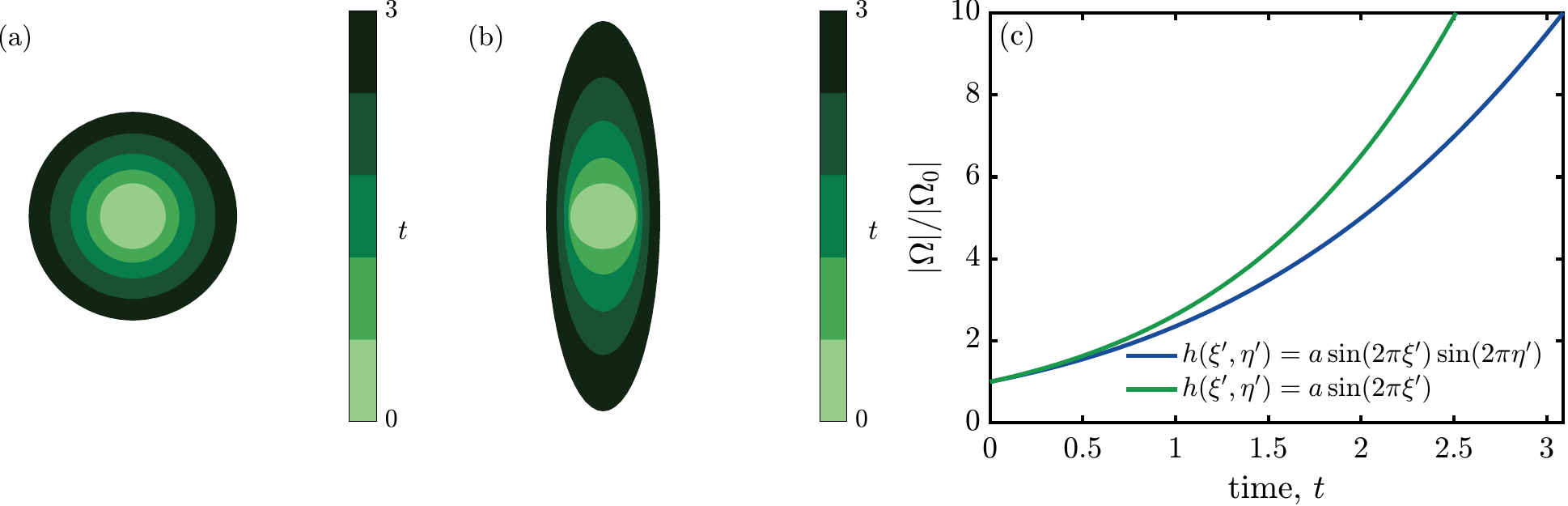}
    \caption{Evolution of the homogenized domain for (a) a two-dimensional wave and (b) a one-dimensional wave. In the former case the domain expands radially while the latter elongates along the direction orthogonal to the wave. (c) Biofilm area over time. The biofilm grows faster over the anisotropic surface compared to the isotropic surface. In each case the domain is initialized as a circle with physical area $|\Omega| = \pi$ and the stress sensitivity is $\lambda = 1$. Each simulation is run until the domain reaches ten times the initial area.}
    \label{fig:anisotropic_moving}
\end{figure*}

These calculations show, for a static geometry, anisotropic surface roughness can enhance growth by reducing stress. Simultaneously, this anisotropy will cause the biofilm to elongate. We assess this time evolution here by numerically simulating the homogenized equations (\ref{eq:u-hom})-(\ref{eq:pbc-hom}). We use the MATLAB package \texttt{surfacefun.m} \cite{fortunato2024high}, which is a fast spectral element solver for PDEs on surfaces. (While this library can handle arbitrary surfaces, we emphasize the homogenized surface is flat.) The domain is discretized using elliptic grid generation \cite{thompson1998handbook} with spectral elements of order $p = 12$, and is remeshed at each time step. We advanced the interface under the homogenized velocity using a fourth-order Runge-Kutta with a time step $\Delta t = 0.05$. In each case we initialize the domain as a circle with area $|\Omega|=\pi$, accounting for the homogenized surface area, and the simulation is run until the area has reached $10\pi$. Figure \ref{fig:anisotropic_moving} shows snapshots of the evolving biofilm for the (a) two-dimensional wave $h(\xi',\eta') = a\sin(2\pi \xi')\sin(2\pi \eta')$ and (b) one-dimensional wave $h(\xi',\eta') = a\sin(2\pi \xi')$. As expected, in the former case the biofilm expands radially while the latter elongates, forming an ellipse-shaped geometry with increasingly high aspect ratio. Panel (c) shows the corresponding domain area over time, where we find the biofilm grows faster over the anisotropic surface compared to the isotropic surface, despite the changing geometry.

\section{Asymptotically optimal surfaces}\label{sec:optimization}

The results of Sections \ref{sec:sphere} and \ref{sec:rough} show the coupled interactions of geometry and stress-sensing can either increase or decrease the growth rate of biofilms of the same size. It is then natural to ask what surfaces, if they exist, minimize or maximize the growth rate $|\dot\Omega|$ for a fixed surface area $|\Omega|$. In this section we show that no such optimal surface can be attained within the class of smooth surfaces of fixed area. In particular, we prove that $\inf_\Omega |\dot\Omega|/|\Omega| = 0$ and $\sup_\Omega |\dot\Omega|/|\Omega| = 1$, which can only be approached by sequences of surfaces with increasingly extreme geometry. Our examples, sharp cones and thin strips, respectively, use the observation that $|\dot\Omega| \propto |\partial\Omega|$ when $\lambda \gg 1$, suggesting the strategy is to minimize/maximize $|\partial\Omega|$ for fixed $|\Omega|$. For concreteness we fix $|\Omega| = \pi$ to be the area of the unit disk, though the arguments hold for arbitrary $|\Omega|$.

\subsection{Minimizing sequence}

Consider first the conical surface $M_\eps$ given by the parameterization
\begin{equation}
    \bx_\eps(\xi, \eta) = \begin{pmatrix} \eps \eta \cos\xi \\ \eps\eta\sin\xi \\ \alpha(\eps) \eta
    \end{pmatrix},\quad (\xi,\eta) \in [0,2\pi) \times [0, \infty),
\end{equation}
where $\alpha(\eps) = \sqrt{1/\eps^2 - \eps^2}$ and $\eps \ll 1$. The associated metric tensor and Jacobian in local coordinates are
\begin{align}
    \bG_\eps(\xi,\eta) &= \begin{pmatrix}
        \eps^2\eta^2 & 0 \\ 0 & 1/\eps^2
    \end{pmatrix},\\
    J_\eps(\xi,\eta) &= \eta.
\end{align}
We define $\Omega_\eps = \{ \bx_\eps(\xi,\eta) : \xi \in [0, 2\pi), \eta \in [0, 1)\}$ so that $|\Omega_\eps| = \pi$ and $|\partial\Omega_\eps| = 2\pi\eps$. In local coordinates, the axisymmetric pressure satisfies
\begin{align}
    -\eps^2\frac{1}{\eta}\frac{\partial}{\partial \eta} \left( \eta \frac{\partial p_\eps}{\partial \eta}\right) = e^{-\lambda p_\eps}, &\quad \eta\in(0,1),\\
    \frac{\partial p_\eps}{\partial \eta} = 0, &\quad \eta = 0,\\
    p_\eps = 0, &\quad \eta = 1.
\end{align}
This is equivalent to the flat solution under the change of variables $p'_\eps = \eps^2 p_\eps$ and $\lambda' = \lambda/\eps^2$. Thus, using Eq. (\ref{eq:uflat}), the kinematic condition (\ref{eq:dX/dt}), and performing a bit of algebra, we find the growth rate is
\begin{equation}
\frac{|\dot\Omega_\eps|}{|\Omega_\eps|} = 8c(\lambda/\eps^2),
\end{equation}
where $c(\lambda') = [(1 + \lambda'/2)^{1/2}-1]/(2\lambda')$ is as in the flat solution. It is straightforward to see $c(\lambda/\eps^2) = O(\eps)$ for fixed $\lambda$, and so this vanishes as $\eps\rightarrow 0$, hence $\inf_{|\Omega|=\pi} |\dot\Omega| = 0$. But $e^{-\lambda p} > 0$ so this cannot be achieved for a domain with non-zero surface area, hence no minimizing surface exists.

\subsection{Maximizing sequence}

Next consider the thin rectangular domain $\Omega_\eps \subset \bbR^2$ given by
\begin{equation}
    \Omega_\eps = \left\{ (\xi,\eta) : |\xi| < \pi/(4\eps), |\eta| < \eps\right\},
\end{equation}
which has $|\Omega_\eps| = \pi$ and $|\partial\Omega_\eps| = \pi/(2\eps) + 2\eps$. Write $p_\eps = q_\eps - (\eta^2-\eps^2)/2$. Then $\Delta q_\eps = 1 - e^{-\lambda p_\eps} \geq 0$, so that, by the maximum principle, $||q_\eps||_{L^\infty(\Omega_\eps)} \leq \eps^2/2$. This implies
\begin{equation}
    \begin{aligned}
    ||p_\eps||_{L^\infty(\Omega_\eps)} &\leq ||q_\eps||_{L^\infty(\Omega_\eps)} + ||(\eta^2-\eps^2)/2||_{L^\infty(\Omega_\eps)} 
    \\ & \leq \eps^2.
    \end{aligned}
\end{equation}
Using the inequalities $e^{-\lambda||p||_{L^\infty(\Omega_\eps)}}|\Omega| \leq |\dot\Omega|$ and $|\dot\Omega| \leq |\Omega|$ and taking $\eps\rightarrow0$, we find $\sup_{|\Omega|=\pi} |\dot\Omega| = \pi$. This can only occur if $p = 0$ everywhere, which, by the maximum principle, is impossible for a domain of non-zero surface area, hence no maximizing surface exists.

The examples above are instantaneous and, for finite $\eps$, do not necessarily imply $\lim_{t\rightarrow\infty}|\dot\Omega_\eps|/|\Omega_\eps|\rightarrow 0$ or $\lim_{t\rightarrow\infty}|\dot\Omega_\eps|/|\Omega_\eps|\rightarrow 1$. For the conical geometry the former limit indeed holds, as we can associate the height of the biofilm $H$ along the cone with $\eps = 1 / H$. The slender geometry is, however, inconclusive as it will necessarily grow thicker. Whether surfaces can be designed such that the time-dependent geometry approaches the infimum/supremum for general initial conditions is an interesting, though highly non-trivial, question.

\section{Discussion}\label{sec:discussion}

Through analysis and simulation of a free boundary model of microbial growth, this work shows collective stress-sensing of microbial communities is closely tied to surface geometry at scales much larger than the individual. The insights given by solutions on the sphere and homogenized solutions for periodic surfaces provide design principles for controlling and inhibiting the spread of biofilms. In particular, anisotropic roughness patterns could be used to steer the direction and speed of growth. We exploited these principles to show the growth rate of the biofilm can be made arbitrarily large, up to the stress-free growth rate, or arbitrarily small. These predictions, especially those on periodically corrugated surfaces, pertain directly to dental biofilms and microbial-resistant plants and may inform control strategies. Moreover, they can be systematically tested experimentally using, for example, 3D printing or soft lithography to manufacture both microstructured and large-scale surfaces \cite{chung2007impact,hall2021bacterial}. Based on dimensional analysis, the dependence on the sensitivity parameter, whose dimensional form may be difficult to control directly, can be assessed by varying the single cell growth rate or the substrate drag coefficient.

Boundary layer arguments show, regardless of the details of the surface, the asymptotic growth rate of stress-sensing biofilms should scale in proportion to $|\partial\Omega|$. For circular biofilms, this yields asymptotically linear growth of the colony radius at long times, an effect that is typically attributed to nutrient depletion \cite{pirt1967kinetic}, though more recently attributed to mechanics \cite{kannan2025spatiotemporal}. In the context of curved surfaces, nutrient diffusion will also be influenced by geometry, and may enhance or counteract growth-inhibiting stresses. The relative contribution of stress to nutrient depletion can be estimated {\em a priori} by comparing the stress-inhibiting length $\propto\lambda^{-1/2}$ (Sec. \ref{sec:model}) to the nutrient penetration depth, typically set by the ratio of nutrient diffusion to the consumption rate \cite{pirt1967kinetic}. When the former is smaller, we expect stress-sensing will dominate the boundary dynamics.

A limitation of the current formulation is its two-dimensionality, for which the the mean curvature is irrelevant. This limitation is particularly acute when the magnitude of the curvature is high and the biofilm grows out of the surface. The growth model naturally generalizes to three dimensions, though the computations become significantly more challenging as both the free boundaries on the surface and normal to it must be determined. Perturbation analysis may be a first approximation in this direction, where the biofilm is assumed to be of uniform thickness and constant pressure in the normal direction. 

The biofilm model can be generalized in a variety of directions. In particular, microbes are typically anisotropic and exhibit collective orientational order that can be characterized as an active nematic fluid. In non-growing active nematics, surface curvature influences the number, motion, and distribution of topological defects \cite{ellis2018curvature,pearce2019geometrical,al2023morphodynamics,nitschke2025active}, and the distribution of these defects will likely influence stress. Moreover, stresses may lead to surface deformation \cite{salbreux2017mechanics,mietke2019self}.

Finally, while we primarily considered structured surfaces, the free-boundary model poses interesting computational challenges for general surfaces. In particular, formulating a method for stable and accurate curve evolution on complex surfaces, such as those deriving from experimental data in the form of a point cloud or mesh, is an interesting direction for future work.

\vspace{0.125in}

\noindent{\bf Acknowledgements.} We thank Dan Fortunato for useful discussions. We used Claude Opus 5.5 (Anthropic) to proofread the manuscript before submission, optimize code, and verify analytical calculations. All text was written by the author. The Flatiron Institute is a division of the Simons Foundation.\\

\noindent{\bf Data Availability.} The data that support the findings of this study were generated by numerical simulations. The source code and parameters used to generate the simulations and corresponding figures are publicly available at \url{https://github.com/scottweady/surface-growth}.

\appendix
\predisplaypenalty=0 

\section{Linear stability of the planar solution}\label{app:stability}

Solutions to the pressure equation (\ref{eq:Lp}) can be expressed explicitly in complex variables \cite{liouville1853equation}. Associating the point $\bx(\eta,\xi) = \eta(\cos\xi,\sin\xi)$ with the complex point $z = \eta e^{\i\xi}$, we have
\begin{equation}
    p(z) = \frac{1}{\lambda}\log\left[\frac{\lambda(1 - |f(z)|^2)^2}{8|f'(z)|^2}\right],\label{eq:liouville}
\end{equation}
where $f(z)$ is holomorphic with $f' \neq 0$ and $|f| < 1$. For the disk, $|z| < H$, we find $f(z) = (1 - 8c)^{1/2} (z/H)$, where $c$ satisfies $8\lambda H^2 c^2 = 1 - 8c$, which is precisely the axisymmetric solution $p_0$ in Eq. (\ref{eq:pflat}).

We consider the perturbed parameterization $\bX(\xi,t) = H(t)(1 + \eps(t)\cos m\xi)\hat\betav$ where $\hat\betav = (\cos\xi,\sin\xi)$. To $O(\eps)$, we find
\begin{equation}
    f(z) = (1 - 8c)^{1/2}\,\frac{z}{H}\left[1 - \eps b \left(\frac{z}{H}\right)^m\right],
\end{equation}
where $b = (1 - 8c)/[1 + 4c(m - 1)]$. Substituting into Eq. (\ref{eq:liouville}) and expanding $p = p_0 + \eps p_1\cos m\xi + O(\eps^2)$, we find
\begin{equation}
    \begin{split}
    p_1(\eta) &= \frac{16c^2H^2}{1 + 4c(m - 1)}\left(\frac{\eta}{H}\right)^m \\
    &\quad\times\left[\frac{(m + 1) - (m - 1)(1 - 8c)(\eta/H)^2}{1 - (1 - 8c)(\eta/H)^2}\right],
    \end{split}
\end{equation}
which satisfies the linearized boundary condition $p_1(H) = -Hp_0'(H) = 4cH^2$. At $O(\eps)$, the kinematic condition (\ref{eq:dX/dt}) then gives $\d\eps/\d t = \sigma\eps$, where ${\sigma = -p_0''(H) - p_1'(H)/H - 4c}$. After some algebra, this gives Eq. (\ref{eq:dispersion}).

\bibliography{refs}

\end{document}